\documentstyle[12pt,epsfig]{article}
\begin{document}
\def\roughly#1
{\raise.3ex\hbox{$#1$\kern-.75em\lower1ex\hbox{$\sim$}}}
\baselineskip0.3in
\vspace{2cm}

\begin{flushright}
JHU-TIPAC-95019\\
hep-ph/9507410\\
to appear in Phys. Lett. B\\
\end{flushright}

{\centerline{\bf \LARGE  A model for the $B\rightarrow X_{s}+
\gamma$ decay }}
{\centerline{\bf \LARGE in the chromomagnetic background}}
\vspace{0.5cm}
{\centerline{\bf \large Alexander Kyatkin}}
\vspace{0.5cm}
{\centerline{\em Department of Physics and Astronomy}}
{\centerline{\em The Johns Hopkins University}}
{\centerline{\em Baltimore MD 21218, USA}}
\vspace{1cm}
\begin{abstract}
\noindent We calculate the shape of the photon spectrum for
$B\rightarrow X_{s}+\gamma$ decay in the presence of a
background chromomagnetic field for the case of SU(2)
 gauge group. The effect of the external field resembles
the Zeeman effect -
the parton model peak is split into two slightly
assymetric peaks, located around
the kinematic endpoint. We investigate the analytic properties of
the spectrum and calculate the total decay rate. This decay may
serve as a model for the investigation of  gluon
condensate effects and their influence on the shape
of the spectrum.
\end{abstract}

\vspace{1cm}

 Recently, there has been  considerable interest in the photon
energy spectrum in the decay $B\rightarrow X_{s}+\gamma$, inspired
 by the first experimental measurement of this decay by the
CLEO collaboration \cite {CLEO}.  Considerable theoretical
progress  has been made by applying  Wilson's operator product
expansion (OPE) and the heavy quark effective theory (HQET) to
the study of the endpoint region of the photon spectrum in this
decay \cite {OPE}. The inclusion of higher order matrix elements
from the OPE smears the parton model spectrum over a region
of the order of $\Delta y\approx \Lambda/m_{b}$, where
$y=2 E_{\gamma}/m_{b}$ ( $E_{\gamma}$ is the $\gamma$-energy and
 $\Lambda$ is the QCD parameter).  The detailed shape of this
nonperturbative spectrum is, however, unknown, because it
depends on the expectation values of
 higher order in $1/m_{b}$ matrix elements. The radiative
corrections give  additional uncertainty to the shape of the
spectrum, because the
 higher order terms in the perturbative expansion are important.
Some attempts have been made to include the higher order
perturbative
terms by resumming the ``renormalon chains" \cite {BLM}. However,
the terms of a very high order in the ``renormalon chain" come
from  large distances and have to be treated separately, as a
condensate \cite {Shifman}.

 In this paper we consider the $B\rightarrow X_{s}+\gamma$ decay
 in the external chromomagnetic background for the case of an
SU(2) gauge group. The possibility that a confining ground state
 in QCD develops a nonzero color background field was discussed a
long time ago \cite {background}.  The decay in the background
field may serve as a model for the investigation of  gluon
condensate effects.  This model allows the nonperturbative
analytical calculation of the photon spectrum and the
detailed investigation of its properties.  The spectrum
depends on one parameter, $p=B^{2}/m_{b}^{2}$, where
$B\sim \Lambda$ is the magnitude of the external field.
We show that the spectrum is nonvanishing  in the
 region $1-\sqrt {p}/2\leq y \leq 1+\sqrt {p}/2$. The shape
of the
 spectrum resembles the shape of spectrum in the case of the
quantum
mechanical Zeeman effect - the initial $\delta$-function peak is
 split into two smeared peaks, located slightly assymetrically
around the kinematic endpoint $y=1$. We also obtain the
analytic expression for the total decay rate and calculate
several moments of the spectrum. First, we review very briefly
the formalism for the theoretical treatment of radiative B decay
 and then calculate
 the photon spectrum in the presence of the background field.

\ The effective Hamiltonian for $B\rightarrow X_{s}+\gamma$
decay may be written as
\begin{equation}
H_{\rm eff}= - {4 G_{\rm F}\over \sqrt {2}} \mid V_{tb}
V_{ts}^{\star} \mid c_{7}^{\rm eff}(m_{b}) \,
O_{7}(m_{b})\ \ ,
\end{equation}
where $O_{7}={e/{16 \pi^2}}\, {\overline {s}}
\,\sigma^{\mu \nu} (m_{b}({1+\gamma_{5}})/2\,+
\,(1-\gamma_{5})/2)\,b \,F_{\mu \nu}$
and  $c_{7}^{\rm eff}(m_{b})$ is the Wilson coefficient
at the scale $\mu\sim m_{b}$ (for the recent review see
\cite {Ali}).
In this paper we assume for  simplicity $m_{s}=0$.

The decay rate can be written in terms of the imaginary
part of a correlator of two local currents \cite {decay}
\begin{equation}
T(v\cdot k)\,=\, - i \int d^{4}x\,{\rm e}^{i(m_{b}v-k)x}
\,\langle B\mid\, T\lbrace J^{\mu}(x),J_{\mu}^{\dagger}(0)
\rbrace\mid B\rangle
\end{equation}
where the current has the form
$$
J^{\mu}(x)\,=\,{\overline b_{v}}(x)\,[\gamma^{\mu},
{\hat k}]\,m_{b} ({1-{\gamma_{5}}\over {2}}) \cdot s(x)
$$
and ${\overline b_{v}}(x)={\rm e}^{i m_{b} v\cdot x}\,
 b(x)$, $k$ is the photon momentum.

The expression for the decay rate is
\begin{equation}
d \Gamma\,=\,{ d^{3}k\over (2\pi)^{3} 2 E_{\gamma}}
{\alpha G_{\rm F}^{2}\over 8\pi^{3}} \mid   V_{tb}
 V_{ts}^{\star} \mid^{2}\,\mid c_{7}^{\rm eff}(m_{b})
\mid^2\,
{\rm Im}\,T(v\cdot k).
\end{equation}
The leading term in $1/m_{b}$ expansion corresponds to
the parton model result and is given by
\begin{eqnarray}
\label{eq:a}
 T(v\cdot k)=\,-i\, m_{b}^{2}\int\,d^{4}x
\,{\rm e}^{i(m_{b}v-k)\cdot x} \nonumber \\
 \langle B\mid {\overline b_{v}}(x) \,[\gamma_{\mu},
{\hat k}] {(1-\gamma_{5})\over 2}\, (i S_{\rm F}(x))
 {(1+\gamma_{5})\over 2}\, [{\hat k},\gamma^{\mu}]
\,b_{v}(0)\mid B\rangle
\end{eqnarray}
where $S_{\rm F}$ is the final $s$-quark free propagator.

This expression leads to the differential decay rate
\begin{equation}
{d\Gamma\over dy}=\, \Gamma_{0}\, \delta (1-y)  \ \ ,
\end{equation}
where $y={2 E_{\gamma}\over m_{b}}$ and
\[
 \Gamma_{0}={\alpha m_{b}^{5} G_{\rm F}^{2}\over 32
\pi^{4}}\,\mid   V_{tb} V_{ts}^{\star} \mid^{2}
\,\mid c_{7}^{\rm eff}(m_{b})\mid^2
\]
 is the parton model total decay rate \cite {decay}.

Now we want to investigate the influence of the
external gluon field on the shape of photon spectrum.
The external field with  field strength of the order
of $\Lambda$  gives a nonperturbative example of how
the gluon condensate corrections to the final state
$s$-quark would affect the spectrum.

 We consider the case of the SU(2) gauge group, which
 allows  analytical calculations. We work in an axially
symmetric anzatz for the background field. The axially
symmetric anzatz may represent the solution of the
 Yang-Mills equations in the strong coupling limit
\cite {gorski}, if we assume that the external source
is axially symmetric, what  represents, probably, the
final  event symmetry of the decay. We neglect also any
space-dependence of the external field.  We nave to note
 the unphysical features of the model, such as the apparent
lack of the Lorentz and gauge invariance due to the presence
of a constant background field. However, we restore,
formally, these symmetries. We take only gauge-independent
contribution to the decay rate and integrate it with
respect to the possible direction of the magnetic field,
restoring the gauge and rotational invariance. To  restore
formally relativistic invariance, we assume that the value
of  parameter of the model corresponds to the
Lorentz invariant condensate value.

Thus, we chose the background SU(2) field configuration as
\begin{equation}
{\bf A}_{0}=0;\ \ \ g {\bf A}_{1}= B {\mbox{\boldmath $
\sigma$}_{1} \over 2};\ \ \  g {\bf A}_{2}=B
{\mbox{\boldmath $\sigma$}_{2}\over 2};\ \ \ {\bf A}_{3} =0
\end{equation}
where $\mbox{\boldmath $\sigma$}_{i}$ are the
{\boldmath $\sigma$} -matrices. This configuration
corresponds to the chromomagnetic field directed along
the third axis
in the configuration and coordinate spaces
\begin{equation}
\label{eq:b}
g {\bf F}_{12}=B^{2}{\mbox{\boldmath $ \sigma$}_{3} \over 2}
\end{equation}

To take into account the effect of this field on the shape
of the spectrum we have to use the $s$-quark propagator in
the presence of the external background. We will find the
analytic expression for the propagator, so  this model may
be considered as a model for the summation of the infinite
series of the gluon condensates insertions into the final
$s$-quark. The expression for the propagator,
\begin{equation}
\label{eq:prop}
G(q)= \,{(\hat {\bf Q}\,)\over \hat {\bf Q^{2}}}=
\,{(\hat {\bf Q})\over  {\bf Q}^2+1/2 i g \sigma_{\mu \nu}
{\bf F}^{\mu \nu}}\ ,\ \ \sigma_{\mu\nu}=1/2
 [\gamma_{\mu},\gamma_{\nu}]
\end{equation}
(where the ${\bf Q}_{\mu}=q_{\mu}+ g {\bf A}_{\mu}$ and
$q=m_{b}\cdot v\,-\,k$) may be found using the Schwinger
 proper time formalism \cite {Schwinger}.
The propagator
 can be written as
\begin{equation}
G(q)\,=\,- i ({\hat q}-1/2\,B
\,(\mbox{\boldmath $ \sigma$} \gamma)_{\perp})
\,\int_{0}^{\infty}\,ds\,
{\rm e}^{i s (q^2-B^2/2)}\,{\rm e}^{i s {\bf D}_{0}},
\end{equation}
where we we use the notation $(a\cdot b)_{\perp} = a_{1}
 b_{1} +a_{2} b_{2}$ and the operator ${\bf  D}_{0}$ is
given by
\[
{\bf D}_{0}=- B (q \mbox{\boldmath $ \sigma$} )_{\perp}
\,+\,B^2/2 \,(\Sigma_{3}\mbox{\boldmath $ \sigma_{3}$} );
 \ \ \ \ \Sigma_{3}=i \gamma_{1}\gamma_{2}.
\]
Using that  ${\bf D}_{0}^2$ is proportional to the unit
operator,
 the result is \cite {Rus}
\begin{eqnarray}
G(q)= - i ({\hat q} - 1/2 B (\mbox{\boldmath $ \sigma$}
 \gamma))_{\perp})\, \int_{0}^{\infty}\,ds
\,{\rm e}^{(i s (q^2 -B^2/2)} \nonumber \\
\lbrace {\rm cos} (s \mid D_{0}\mid) +
{i {\bf D}_{0} \over \mid D_{0}\mid} {\rm sin} (s
 \mid D_{0}\mid) \rbrace\ \ ,
\end{eqnarray}
where $\mid D_{0}\mid =B {\sqrt {q_{\perp}^2+B^2/4}}$.

In the space of the SU(2) gauge group, the propagator may
be written as
\begin{equation}
\label{eq:c}
G(q)=C_{0}\,+\,C_{1}\mbox{\boldmath $ \sigma$}_{1} \,+
\,C_{2}\mbox{\boldmath $ \sigma$}_{2} \,
+\,C_{3}\mbox{\boldmath $ \sigma$}_{3} ,
\end{equation}
where $C_{i}$ are the functions of the moment.

We have to plug  this propagator into  the expression
(\ref{eq:a}) for the $T$ operator and take the heavy quark
matrix elements. Because the result has to be
gauge invariant,
 we assume that only the $C_{0}$ term (which is  gauge
independent) gives  contribution to  $T$.
Moreover, we will
integrate the imaginary part of $T$ with
respect to the
possible direction of the magnetic field, restoring,
therefore,  the rotational invariance.

The expression for the $C_{0}$ is
\begin{equation}
C_{0}={{\hat q} (q^2\,-\,B^2/2)\,-\,B^2/2
(q\gamma)_{\perp}\over (q^2-B^2/2)^2-B^2 (q^2_{\perp}+B^2/4)}.
\end{equation}

The final expression for  $T$ has the following form
\begin{equation}
T=\, - 4 m_{b}^{4}\,y^{2}\,{(1-y-p/2+p/4 \,y \,{\rm sin}^{2}
\Theta)\over (1-p/4)(y-y_{1})(y-y_{2})+p/4 \,y^{2}\,
{\rm cos}^{2}\Theta}
\end{equation}
where we introduce the parameter $p=B^{2}/m_{b}^{2}$,
$\Theta$ is the angle between the space direction of the
 magnetic field and the photon 3-momentum ${\bf k}$. Here
\[
y_{1,2}=1+{(-p\mp 2{\sqrt p})\over (4-p)}
\]
The imaginary part of $T$ is nonzero for $y_{1}\leq y
\leq y_{2}$. For  small $p$ it corresponds to a cut
(the branch singularity), located at
\begin{equation}
\label{eq:d}
1-{{\sqrt p}\over 2}\leq y \leq 1+{{\sqrt p}\over 2}
\end{equation}

We take the imaginary part of $T$ and integrate it with
respect to the angle $\Theta$. The regularization
prescription $q^{2}\rightarrow q^{2}+i\epsilon$
(where $q=m_{b} v\,-\,k$) is used. Also we use the
prescription
\[
{\rm Im} {1\over x\pm i\epsilon}=\mp \pi \,\delta (x).
\]

The result for the differential decay rate is
\begin{eqnarray}
\label{eq:e}
{d\Gamma\over dy}&=&\cases{\displaystyle
{{\Gamma_{0}\over
{\sqrt p}} \,({y^{2}(1-y-p/2+p/4
\,y)\over {\sqrt {(1-p/4)(y-y_{1})(y_{2}-y)}}}
\,-\,y\,
{\sqrt {(1-p/4)
(y-y_{1})(y_{2}-y)}})}&\cr
\ \ \ \ \ \ \ \ \ \ \ for\ $$ y_{1}\leq y
\leq
{y_{1}+y_{2}\over 2}$$ \cr
{}~&~\cr
\displaystyle{{\Gamma_{0}\over {\sqrt p}}
\,(-\,{y^{2}
(1-y-p/2+p/4 \,y)\over {\sqrt {(1-p/4)(y-y_{1})
(y_{2}-y)}}}\,+\,y\, {\sqrt {(1-p/4)
(y-y_{1})(y_{2}-y)}})}&\cr
\ \ \ \ \ \ \ \ \ \ \ for\   $${y_{1}+y_{2}
\over 2}
\leq y \leq y_{2}$$ \cr
}
\end{eqnarray}
The shape of the spectrum is shown in Fig.(1).
The spectrum has two peaks, located at $y_{1,2}\approx 1\mp
{{\sqrt p}\over 2}$, smeared over the region between this
 points. Thus, the effect from the chromomagnetic field
resembles the Zeeman effect - the parton model single peak,
located at the kinematic endpoint $y=1$, is split into two
peaks. The right peak corresponds to the final $s$-quark
with the polarization directed along the magnetic
field, while the left peak corresponds to the $s$-quark
 with the opposite
 polarization. The spectrum is negative in the region
$(1-p/2)\roughly{<} y \roughly{<} (1-p/4)$ and has a
discontinuity
 at $y\approx 1-p/4$.

	\centerline{\epsfig{file=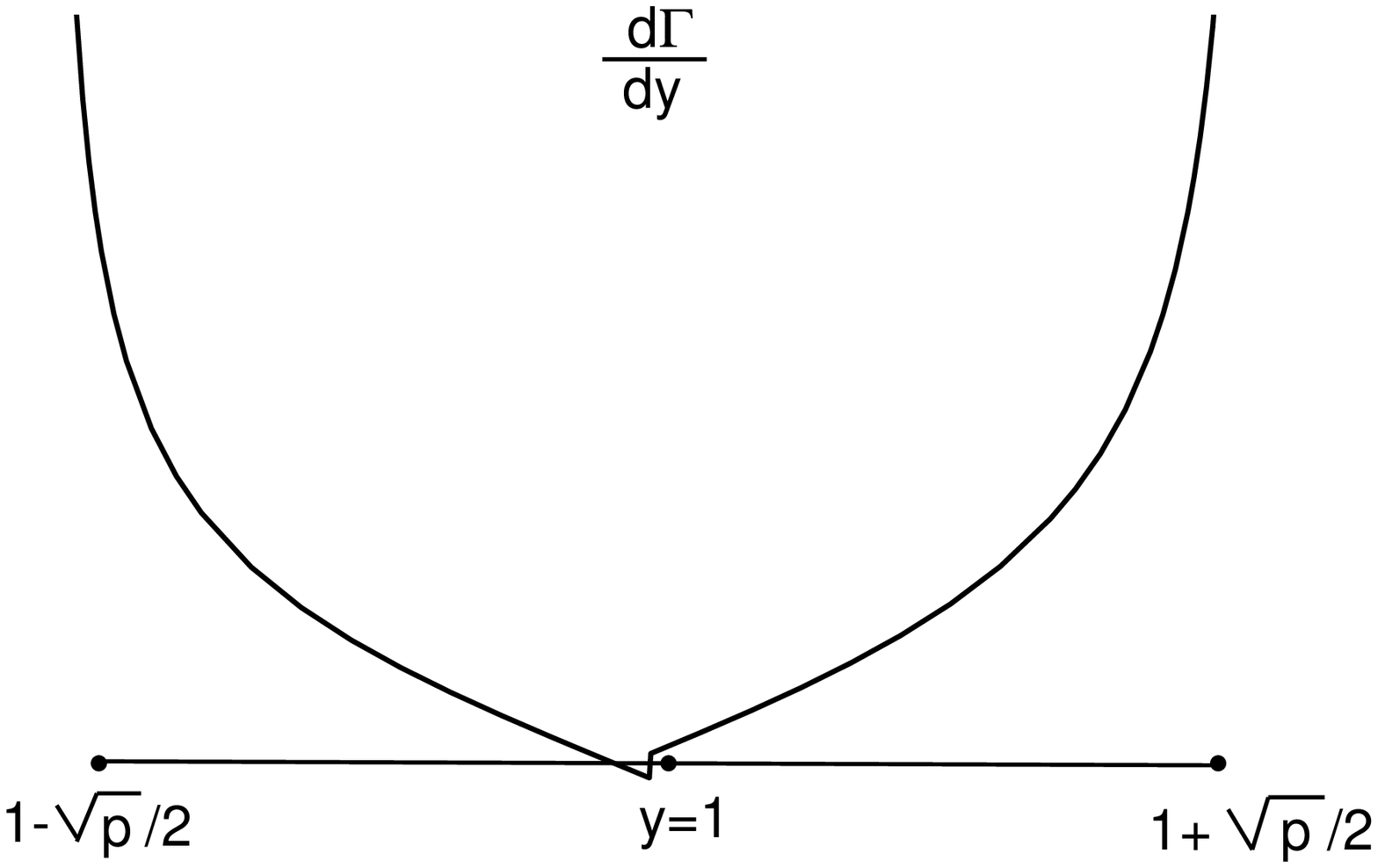,height=8cm}}
 	\centerline{Fig.1: The differential decay rate
$d\Gamma/dy$ }
         \centerline{in the presence of the background
field.}

We suppose that this negative ``gap"
 and discontinuity are spurious and related to the presence
of the constant background field in this model. For a
gauge-dependent background field $C_{i}$ ($i=1,2,3)$ terms
in propagator (\ref{eq:c}) give additional contribution to
the $T$ operator (\ref{eq:a}) and may cure these defects.
Moreover, the motion of the $b$-quark inside  the B-meson
\cite {motion} and the radiative corrections must smear the
spectrum and ``raise" it in this region.

We calculate  the total decay rate in this model. The result
is
\begin{equation}
\label{eq:f}
\Gamma\,=\, \Gamma_{0} {8\,(16-16p+7p^2-p^3)\over
(4-p)^{7/2}}=\Gamma_{0} \,(1-{p\over 8}+{7\over 128}
p^2\,+\,O(p^{3}))
\end{equation}
 We note that the contribution from the right peak to the
total decay rate is slightly larger then the contribution
from the left peak; the difference is of the order of
$O({\sqrt p})$. This assymetry  arises because the $s$-quark
polarization along the magnetic field is  more
preferable
energetically than the opposite polarization. We suppose
that assymetry of the spectrum with respect to the
kinematic
endpoint $y=1$ is  a general feature of the spectrum,
related
to the presence of $\sigma_{\mu \nu} {\bf F}^{\mu \nu}$ term
in  propagator (\ref{eq:prop}).

We calculate also the  moments of the photon energy spectrum
in this model. We define the moments as
\[
{\overline a}_{k}=\int (y-s)^{k} {1\over \Gamma}
{d\Gamma\over dy}\,dy,
\]
where we normalize the differential decay rate on the total
decay rate (\ref{eq:f}) and define the moments around the
reference point $y=s\approx\,1+p/6$. For such a choice
\[
{\overline a}_{0}=1,\ \ \ \ \ {\overline a}_{1}=0
\]
The next two moments are
\begin{equation}
\label{eq:g}
{\overline a}_{2}=p/6,\ \ \ \ \ {\overline a}_{3}=
-{2\,p^2\over 15}.
\end{equation}

 Even for the case of SU(2) gauge group it is helpful to
make some estimate for the value of the parameter $p$.
Restoring formally the relativistic invariance, we assume
that the value of parameter $p^2$ corresponds to the
condensate value
\[
p^2 \sim {{\langle 0\mid \alpha \,Tr\,{\bf F}_{\mu\nu}
{\bf F}^{\mu\nu}\mid 0 \rangle}\over m_{b}^4}
\]
Because the condensate value is of the order of
$\Lambda^4/m_{b}^4$, the value of the parameter $p$ is
$p\approx \Lambda^2/m_{b}^2$. Thus, according to the
equation (\ref{eq:f}),  the expansion around $p=0$
resembles the $1/m_{b}$ expansion \cite {motion}. The
estimate for the moment $a_{2}={\overline a}_{2}\,
m_{b}^{2}/\Lambda^2$ is $a_{2}\sim 0.2$
and $a_{3}={\overline a}_{3}\, m_{b}^{3}/\Lambda^3\,
\sim\, -0.01$. The recent estimates for these moments
are $a_{2}\sim 0.7$, $a_{3}\sim -0.3$ (see \cite {dikeman}
for the review), so the value of second moment
and the sign of the third moment are consistent with
 these estimates.

 We want to note that the absence of $\sqrt {p}\approx
\Lambda/m_{b}$ term in the total decay rate (\ref{eq:f})
is in agreement with Luke's theorem \cite {Luke}, and the
 size of the smeared region of the spectrum $\sqrt
{p}\approx
\Lambda/m_{b}$ (what may be seen from (\ref{eq:d})) is
consistent with the estimates in other models
\cite {motion}.

 In summary, we have calculated the shape of the photon
spectrum in the chromomagnetic SU(2) field.  This model may
 be considered as a model for the summation of the infinite
series of the gluon  condensates. The effect of the external
field resembles the Zeeman effect. We investigated the
analytical properties of the spectrum and calculated the
total decay rate.
In a more rigorous model we would take into consideration
the fluctuation of the magnitude and the direction of the
background field. We expect that these fluctuations lead to
overlapping of the peaks in Fig. (1).  Corrections due to
motion of the $b$-quark inside the B-meson \cite {motion}
and radiative corrections should smear the spectrum further,
leading to the broad peak, observed in the CLEO
experiment \cite {CLEO}.

\vspace{1cm}

The author is grateful to A. Falk for useful discussions
and critical remarks. This work was supported by the
National Science Foundation under Grant No.~PHY-9404057.

\end{document}